\documentclass[]{emulateapj}
\usepackage{placeins}
\usepackage{graphicx}
\usepackage{amssymb, amsmath}
\newcommand{\rns}{\rho_{\rm sat}}
\newcommand{\Ms}{M_{\odot}}
\usepackage{float}
\usepackage{color}

\begin{document}

\title{From Neutron Star Observables to the Equation of State.\\II. Bayesian Inference of Equation of State Pressures}
\author{Carolyn A. Raithel\altaffilmark{1}, Feryal \"Ozel\altaffilmark{1}$^{,}$\altaffilmark{2}, Dimitrios Psaltis\altaffilmark{1}$^{,}$\altaffilmark{3}}

\altaffiltext{1}{Department of Astronomy and Steward Observatory, University of Arizona, 933 N. Cherry Avenue, Tucson, Arizona 85721, USA}
\altaffiltext{2}{Guggenheim Fellow}
\altaffiltext{3}{Radcliffe Institute for Advanced Study, Harvard University, Cambridge, MA 02138, USA}

\begin{abstract}
One of the key goals of observing neutron stars is to infer the equation of state (EoS) of the cold, ultradense matter in their interiors. We present here a Bayesian statistical method of inferring the pressures at five fixed densities, from a sample of mock neutron star masses and radii. We show that while five polytropic segments are needed for maximum flexibility in the absence of any prior knowledge of the EoS, regularizers are also necessary to ensure that simple underlying EoS are not over-parametrized. For ideal data with small measurement uncertainties, we show that the pressure at roughly twice the nuclear saturation density, $\rns$, can be inferred to within 0.3~dex for many realizations of potential sources of uncertainties. The pressures of more complicated EoS with significant phase transitions can also be inferred to within $\sim$30\%. We also find that marginalizing the multi-dimensional parameter space of pressure to infer a mass-radius relation can lead to biases of nearly 1~km in radius, towards larger radii. Using the full, five-dimensional posterior likelihoods avoids this bias.
\end{abstract}

\maketitle

\section{Introduction}
Observations of neutron stars provide a unique way to study the equation of state (EoS) of cold, ultradense matter at high densities. Many theoretical EoS have been proposed, incorporating a wide variety of physics and calculation methods. The sample of proposed EoS ranges from purely nucleonic models (e.g., \citealt{Baym1971, Friedman1981, Akmal1998, Douchin2001}), to models including hyperons (e.g., \citealt{Balberg1997}), pion condensates (e.g., \citealt{Pandharipande1975}), or kaon condensates (e.g., \citealt{Kaplan1986}). A number of newer studies incorporate the effects of the additional quark degrees of freedom expected at high densities, either from phenomenological models or from the early results of lattice QCD (e.g., \citealt{Alford2005, Alford2013, Kojo2015}).

While nuclear physics experiments can be used to inform and constrain these models at low densities, experimental data become sparse even in the vicinity of the nuclear saturation density, $\rns~\sim~2.7~\times~10^{14}$~g~cm$^{-3}$,  for neutron-rich matter (see \citealt{Lattimer2012}). Neutron stars, with central densities up to $8~\rns$, offer a way to constrain the EoS at higher densities. Any particular EoS predicts a unique combination of allowed neutron star masses and radii. Observations of these quantities can, therefore, be compared to the predictions of various EoS models to provide constraints. 

The masses of numerous neutron stars have already been measured, of which at least a dozen have simultaneous radii measurements (\citealt{Guillot2013, Guillot2014, Heinke2014, Nattila2015, Bogdanov2016, Ozel2016a}; for a recent review, see \citealt{Ozel2016}). Some of these measurements already place stringent constraints on the EoS:  for example, the observations of two, roughly 2~$\Ms$ neutron stars (\citealt{Demorest2010}, see \citealt{Fonseca2016} for the revised mass; \citealt{Antoniadis2013}) imply that the EoS must be relatively stiff in order to support the mass against gravitational collapse. However, this approach of using observations to constrain individual EoS models is inherently limited in scope:  it is possible that the current sample of proposed EoS does not probe the full range of physical possibilities.

One alternate approach to constraining individual EoS models is to infer, instead, their functional form directly from the observations, by exploiting the one-to-one mapping of the EoS to the mass-radius curve. It has been shown that it is formally possible to invert that one-to-one mapping to recover the EoS, given sufficient mass and radius observations \citep{Lindblom1992}. Such an inversion requires observations that completely populate the mass-radius curve. However, because there is no known mechanism for producing neutron stars below $\sim 1 \Ms$ and none have been observed, performing a full inversion is observationally unrealistic.

A second approach is to infer a simpler functional form, using a parametric EoS. This requires fewer observations, allows for a more limited data range, and, at the same time, allows for different types of data, such as the moment of inertia, to be combined. Many parametrizations have been proposed. It has been shown, for example, that the EoS can be written as a spectral expansion in terms of the enthalpy \citep{Lindblom2012, Lindblom2014}. It has also been shown that the EoS can be represented as a discrete number of segments that are piecewise polytropic or linear  \citep{Read2009, Ozel2009, Hebeler2010, Steiner2016}. In \citet{Raithel2016}, we performed an optimization of the segmented, parametric EoS and found that using \textit{five} piecewise polytropic segments above the nuclear saturation density was necessary to reproduce the mass, radius, and moment of inertia for a variety of diverse EoS to within the expected uncertainties of next-generation experiments, i.e., to within 0.5~km, 0.1~$\Ms$, and $\sim 10 \%$, respectively.

Given a parametric EoS, the parameters can be determined from a Bayesian inference method. Such methods have been previously developed in \citet{Steiner2010, Steiner2016} and \citet{Ozel2016a}. While a Bayesian method has the potential of being robust, important statistical considerations have yet to be explored. For example, it is not obvious what types of prior distributions should be used; the effects of measurement uncertainties have not yet been quantified; and biases in currently-used methods of marginalization persist. These uncertainties are particularly important when there is an absence of data at low masses to constrain the solution.

In this paper, we perform the natural next step to \citet{Raithel2016}, i.e., using Bayesian statistical techniques to infer the pressures of our optimized five-polytrope parameterization given samples of simulated astrophysical measurements. We review our optimal parametrization in $\S$\ref{sec:paramEoS} and introduce the Bayesian inference method in $\S$\ref{sec:methods}. In $\S$\ref{sec:data}, we test the Bayesian inference on a number of equations of state, ranging from simple polytropes to complex EoS with significant phase transitions.  

We find that, if one wants to be fully agnostic about the physics of the EoS at high densities, the full five-polytrope parametrization must be used. This is the only way to ensure that complex structure and/or significant phase transitions are allowed in the inferred EoS. However, if the true EoS is relatively simple or even a single polytrope, the five-polytrope model will lead to over-parametrization. We find that a regularizer can help reduce the effects of the over-parametrization, while still allowing the complexity that is the benefit of the five-polytrope model.

Finally, in either case, it is important to use the full five- (or three-) dimensional posterior likelihoods rather than a marginalization, which, as we demonstrate in $\S$\ref{sec:marg}, can introduce biases as large as 1~km.

\section{The Optimal Parametric Equation of State}
\label{sec:paramEoS}
In order to infer the functional form of the dense-matter equation of state, we use a parametric EoS composed of piecewise polytropes. As we showed in \citet{Raithel2016}, a parametric EoS with five polytropic segments is optimal for reproducing the observable properties of neutron stars for a diverse set of EoS, to within expected observational uncertainties. 

We begin the parametric EoS at the nuclear saturation density, $\rns$, below which we assume a low-density EoS. We space the five polytropic segments evenly in the logarithm of density, so that the fixed densities separating the polytropes are at 1, 1.4, 2.2, 3.3, 4.9, and 7.4~$\times\rns$. The EoS across each segment is given by
\begin{equation}
P(\rho) = K_i \rho^{\Gamma_i}, \quad  \rho_{i-1} < \rho < \rho_i,
\end{equation}
where the constant $K_i$ and the polytropic index, $\Gamma_i$ are determined by the pressure and density of the previous fiducial point, i.e., 
\begin{equation}
K_i = \frac{P_{i-1}}{\rho_{i-1}^{\Gamma_i}} = \frac{P_i}{\rho_i^{\Gamma_i}}
\end{equation}
and
\begin{equation}
\Gamma_i = \frac{\log_{10}(P_i/P_{i-1})}{\log_{10}(\rho_i/\rho_{i-1})}.
\end{equation}

The above EoS can be mapped to a mass-radius curve using the Tolman-Oppenheimer-Volkoff (TOV) equations. The TOV equations relate the pressure, $P$, and enclosed mass, $M$, of the star as a function of radius, according to
\begin{equation}
\frac{\text{d}P}{\text{d}r} = - \frac{G}{c^2} \frac{(\epsilon + P)(M + 4\pi r^3 P/c^2)}{r^2 - 2GMr/c^2}
\label{eq:dpdr}
\end{equation}
and
\begin{equation}
\frac{\text{d}M}{\text{d}r} = \frac{4 \pi r^2 \epsilon}{c^2},
\label{eq:dmdr}
\end{equation}
where the energy density, $\epsilon$, is given by
\begin{equation}
\label{eq:thermo}
d\frac{\epsilon}{\rho} = -P d\frac{1}{\rho}.
\end{equation}

The TOV equations provide a one-to-one mapping of the EoS to mass-radius space. The goal of this paper is to demonstrate the reverse process: the inference of our parametric EoS, given simultaneous  observations of masses and radii.

\section{Bayesian Inference of EoS Pressures}
\label{sec:methods}
In order to infer the pressures of our parametric EoS, we follow the Bayesian approach of \citet{Ozel2016a}, which we recreate below. (For a similar analysis, see \citealt{Steiner2010}). The pressures of interest are those at our five fiducial densities, which, as described in $\S$~\ref{sec:paramEoS}, completely determine our piecewise-polytropic EoS.

The posterior that a particular realization of our parametric EoS correctly describes a set of data can be written as
\begin{equation}
P( \mathrm{EoS | data}) = P(P_1, P_2, P_3, P_4, P_5 | \mathrm{data}).
\end{equation}
By Bayes' theorem, we can rewrite this as
\begin{multline}
\label{eq:fullL}
P(P_1, P_2, P_3, P_4, P_5 | \mathrm{data}) = \\ 
C P(\mathrm{data} | P_1, ..., P_5 ) \times
 P_p(P_1, ..., P_5),
\end{multline}
where $C$ is a normalization constant, $P_p(P_1, ... , P_5)$ is the prior on the set of five pressures, and
\begin{equation}
\label{eq:alldata}
P(\mathrm{data} | P_1, ..., P_5 ) = \displaystyle \prod_{i=1}^N P_i(M_i, R_i | P_1, ..., P_5 ) 
\end{equation}
is the likelihood of a particular realization of $N$ total mass-radius observations, given a set of EoS parameters. 

To calculate the likelihood of observing a particular value of $(M, R)$ given an EoS, we compute the probability that the observation is consistent with each point along the predicted M-R curve, and then take the maximum likelihood. That is, we calculate
\begin{multline}
\label{eq:ptOnLine}
P_i (M_i, R_i | P_1, ..., P_5 ) = 
 P_{\mathrm{max}}(M_i, R_i | P_1, ..., P_5, \rho_c) , 
\end{multline}
where we have used the central density, $\rho_c$, to parametrize the mass-radius curve. In the limit of small errors in either $M$ or $R$, this method is equivalent to taking the ``closest approach" of the curve to the data point.

Finally, the likelihood that an observation of $(M_i, R_i)$ is consistent with a point on the mass-radius curve is given by
\begin{multline}
\label{eq:obs_prob}
P_i(M_i, R_i | P_1, ..., P_5, \rho_c ) = \frac{1}{2 \pi \sigma_{R_i} \sigma_{M_i}} \\ \exp \Bigl\{  -\frac{ [M_i - M_{\mathrm{EoS}}(\rho_c)]^2}{2 \sigma_{M_i}^2} - \frac{[R_i - R_{\mathrm{EoS}}(\rho_c)]^2}{2 \sigma_{R_i}^2} \Bigr\},
\end{multline}
where $\sigma_{R_i}$ and $\sigma_{M_i}$ are the measurement uncertainties associated with the radius and mass, respectively. Here, $R_{\rm EoS}(\rho_c)$ and $M_{\rm EoS}(\rho_c)$ are the radius and mass predicted by the set of pressures, $(P_1, ..., P_5)$, that comprise our parametrized EoS for a particular central density, $\rho_c$. In order to populate the five-dimensional posterior of eq.~(\ref{eq:fullL}), we use Markov-Chain Monte Carlo simulations following the Metropolis-Hastings algorithm.

\subsection{Priors on the pressures}
\label{sec:labpriors}

For the priors on $(P_1, ..., P_5)$ in eq.~(\ref{eq:fullL}), we employ constraints from physical principles, laboratory nuclear physics experiments, and astrophysical observations. Specifically, we require that: \\
(1.) The EoS be microscopically stable, i.e.,
\begin{equation}
P_i~\le~P_{i+1}.
\end{equation} 
(2.) The EoS remain causal between the fiducial densities, i.e.,
\begin{equation}
\frac{dP}{d\epsilon} = \frac{c_s^2}{c^2} \le 1, 
\end{equation}
where $c_s$ is the local sound speed.  \\
(3.) Each EoS produce a neutron star with a mass of at least 1.97~$\Ms$, in order to be within $1~\sigma$ of the mass measurement of the most massive neutron stars \citep{Antoniadis2013, Fonseca2016}. \\
(4.) Pressures $P_1 \ge 3.60$~MeV/fm$^3$ and $P_2 \ge 11.70$~MeV/fm$^{3}$, in order to be consistent with nuclear physics experiments.

Constraint (4) provides lower limits on the pressures at the first two fiducial densities. As noted in \citet{Ozel2016a}, important constraints on the EoS in the density regime around $\rns$ are obtained through nucleon-nucleon scattering experiments at energies below 350~MeV and from the properties of light nuclei. Results from such experiments can be extended by assuming two- and three-body potentials at densities near $\rns$ \citep{Akmal1998}. However, the interactions at higher densities cannot be written in terms of static few-body potentials. We, therefore, impose this nuclear physics constraint only on the pressures at our two lowest fiducial densities: $\rho_1 = 1.4~\rns$ and $\rho_2 =  2.2~\rns$. Because the three-nucleon interaction is always repulsive, the most model-independent lower limit uses only the two-nucleon interaction. Using the Argonne AV8 two-nucleon pressure as calculated in \citet{Gandolfi2014}, we find $P(\rho_1)$=3.60~MeV/fm$^3$ and $P(\rho_2)$=11.70~MeV/fm$^3$. The AV8 potential is a simplified version of the Argonne AV18 potential \citep{Wiringa1995}; however, as noted in \citet{Ozel2016a}, the two-nucleon interaction pressures are approximately the same for either version of the potential.

\subsection{Regularizers}
\begin{figure}[ht]
\centering
\includegraphics[width=0.45\textwidth]{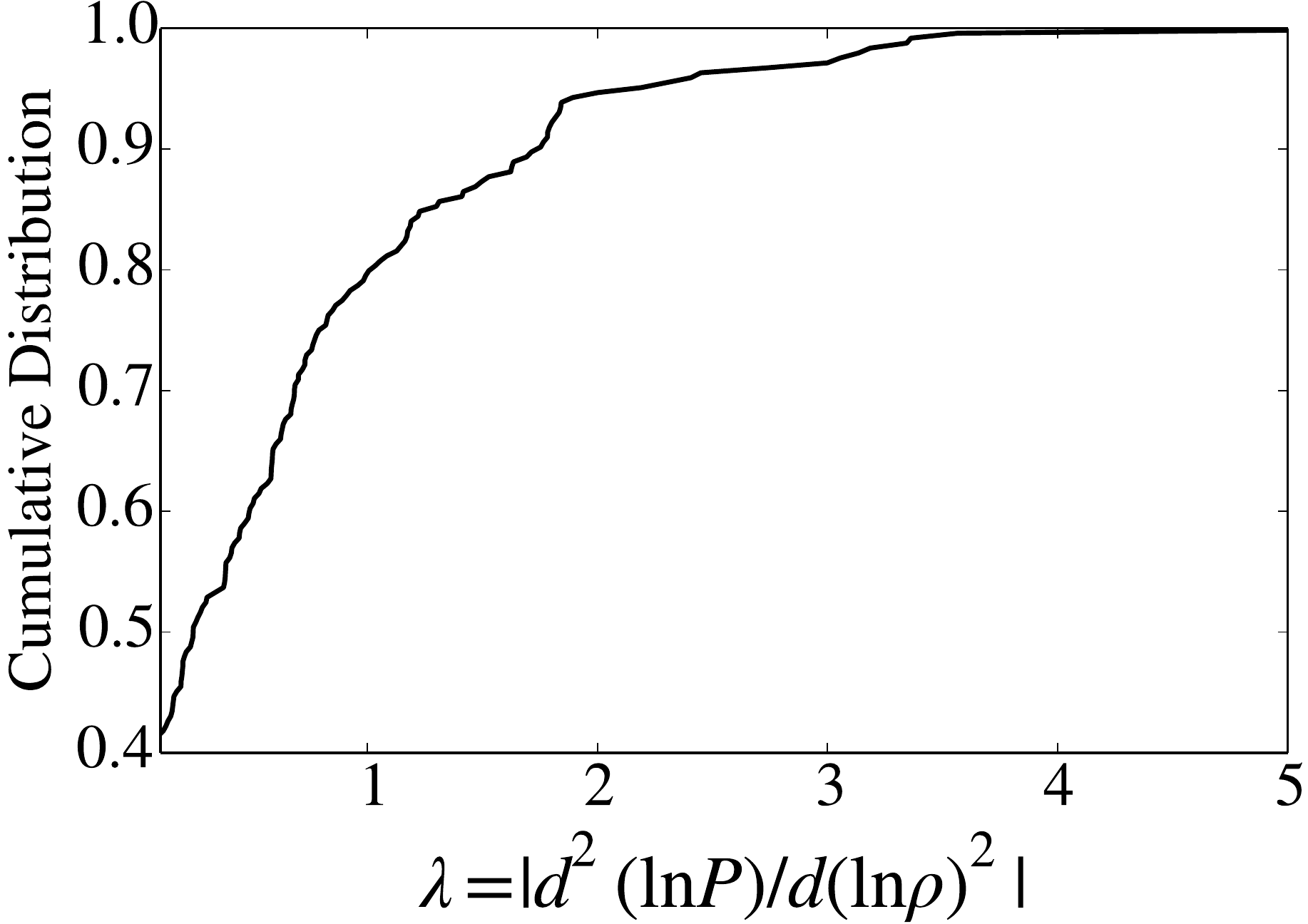}
\caption{\label{fig:cumD2} Cumulative distribution of the second logarithmic derivative in pressure at our five fiducial densities for a sample of 49 EoS taken from the literature. The majority of second derivatives are $\lesssim$~2. We, therefore, take a conservative value of $\lambda=8$ in our Gaussian regularizer for this second derivative.}
\end{figure}

We also include a regularizer in our prior distributions, in order to reduce the tendency of our model to over-parameterize simple EoS. The regularizer, $\xi$, is Gaussian over the second logarithmic derivative of the EoS, i.e., 
\begin{equation}
\xi = \exp \left[ - \frac{( d^2 (\ln{P})/d (\ln{\rho})^2 )^2} {2 \lambda^2} \right],
\end{equation}
where $\lambda$ is the characteristic scale. We determine a suitable value for $\lambda$ by calculating $d^2 (\ln{P})/d (\ln{\rho})^2$ at our fiducial densities for a sample of 49 proposed EoS. This sample of EoS was compiled from the literature in order to incorporate a wide variety of physics and calculation methods, as in \citet{Read2009}. The cumulative distribution of second logarithmic derivatives for this sample are shown in Fig.~\ref{fig:cumD2}. From this cumulative distribution, we find that 95\% of the derivatives are $\lesssim~2$. We, therefore, use a characteristic scale of 4~$\times$ this value, resulting in $\lambda=8$ as our Gaussian regularizer. Such a regularizer will apply, at most, a penalty of $\sim$~3\% to the likelihood that we would calculate for an EoS that has second derivatives that occur in our sample of physically-motivated EoS.

\begin{figure*}[ht]
\centering
\includegraphics[width=0.95 \textwidth]{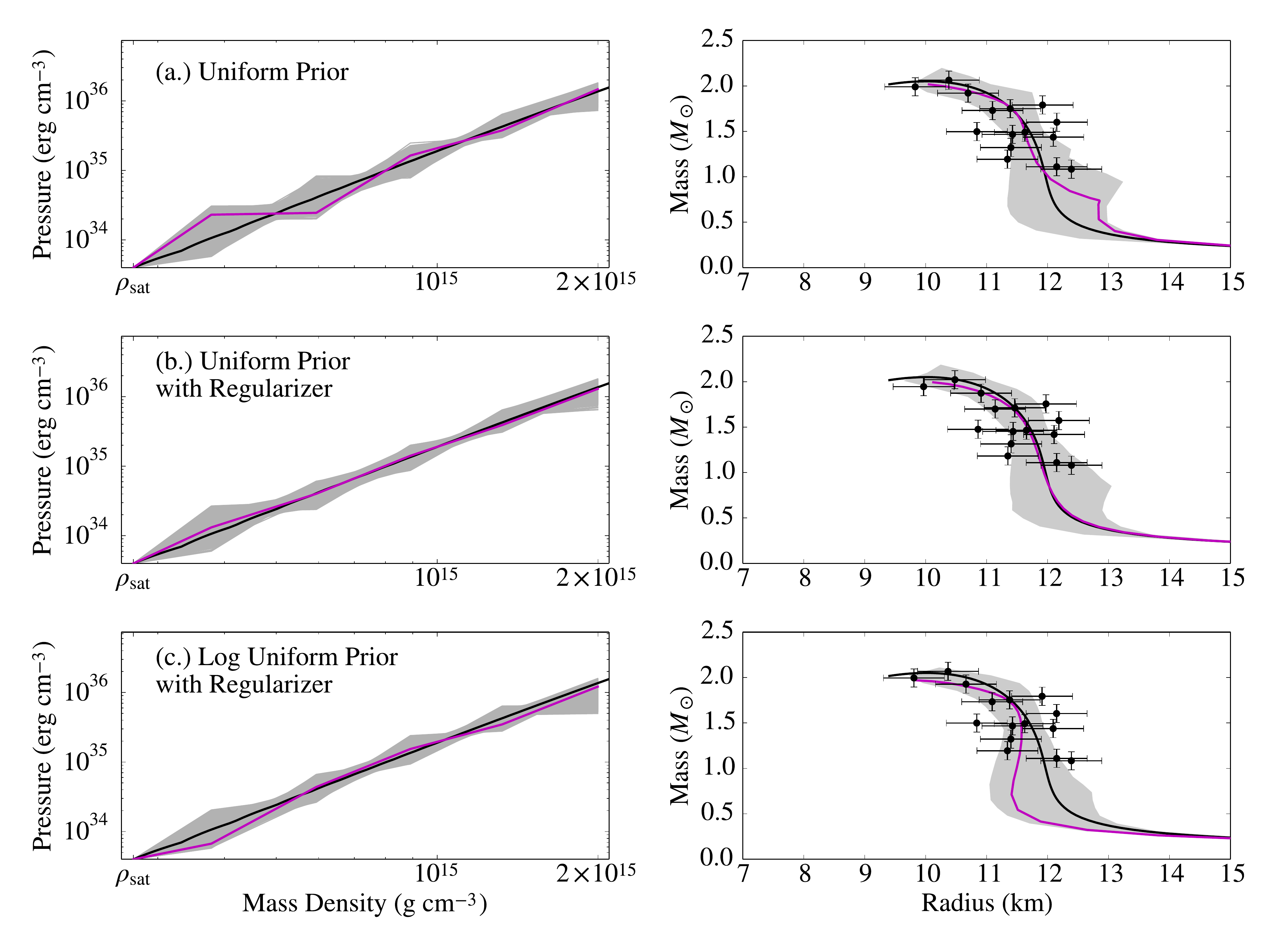}
\caption{\label{fig:uniln}  \textit{(Top)} Inferred equation of state and mass-radius curve from a sample of mock data, assuming a \textit{uniform} prior distribution of pressures. The mock data are drawn from the nucleonic EoS SLy \citep{Douchin2001} and are dithered with Gaussian noise corresponding to $\sigma_M = 0.1 \Ms$, $\sigma_R = 0.5$~km. The actual curves for SLy are shown in black. The magenta curve represents the most likely EoS inferred via our Bayesian method. The 68\% credibility region is shown in gray. \textit{(Middle)} Identical to top panel, but with our Gaussian regularizer included in the inversion. \textit{(Bottom)} Identical data to the top two panels, but assuming a prior distribution that is \textit{uniform in the logarithm} of pressure and including a Gaussian regularizer. Assuming a uniform distribution leads to a preference towards high pressures in the regions where there are few data to constrain the inversion, while assuming that the pressures are distributed uniformly in the logarithm leads to a preference towards lower pressures. Including the Gaussian regularizer reduces the sensitivity to the choice of prior. }
\end{figure*}
\begin{figure}[ht]
\centering
\includegraphics[width=0.45\textwidth]{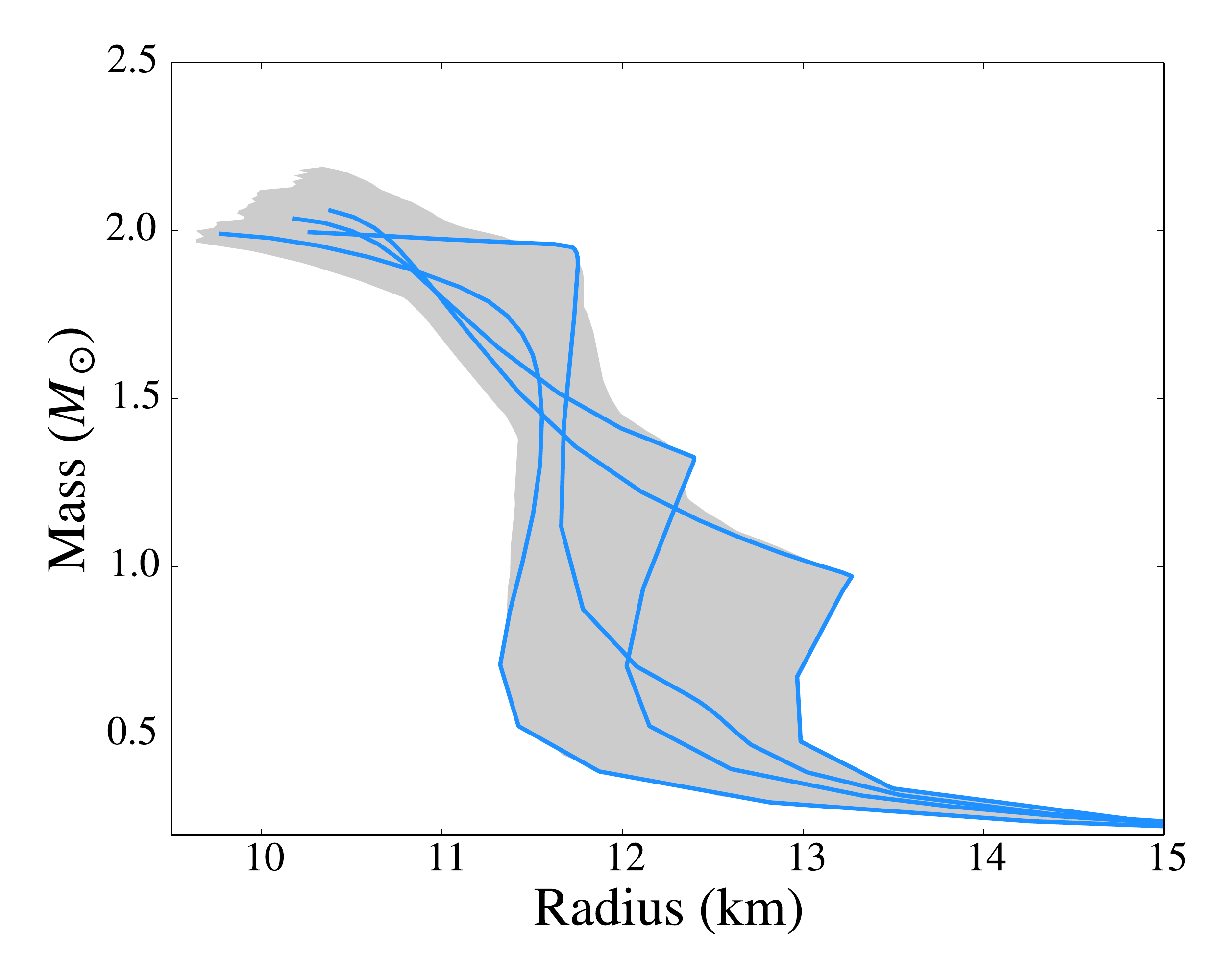}
\caption{\label{fig:indivMR} Individual mass-radius curves contributing to the shape of the 68\% credibility region in the top panel Fig.~\ref{fig:uniln}. A few individual curves are shown here to emphasize the fact that not all curves that can be drawn through this region will actually have likelihoods within the 68\% interval. }
\end{figure}
\begin{figure*}
\centering
\includegraphics[width=\textwidth]{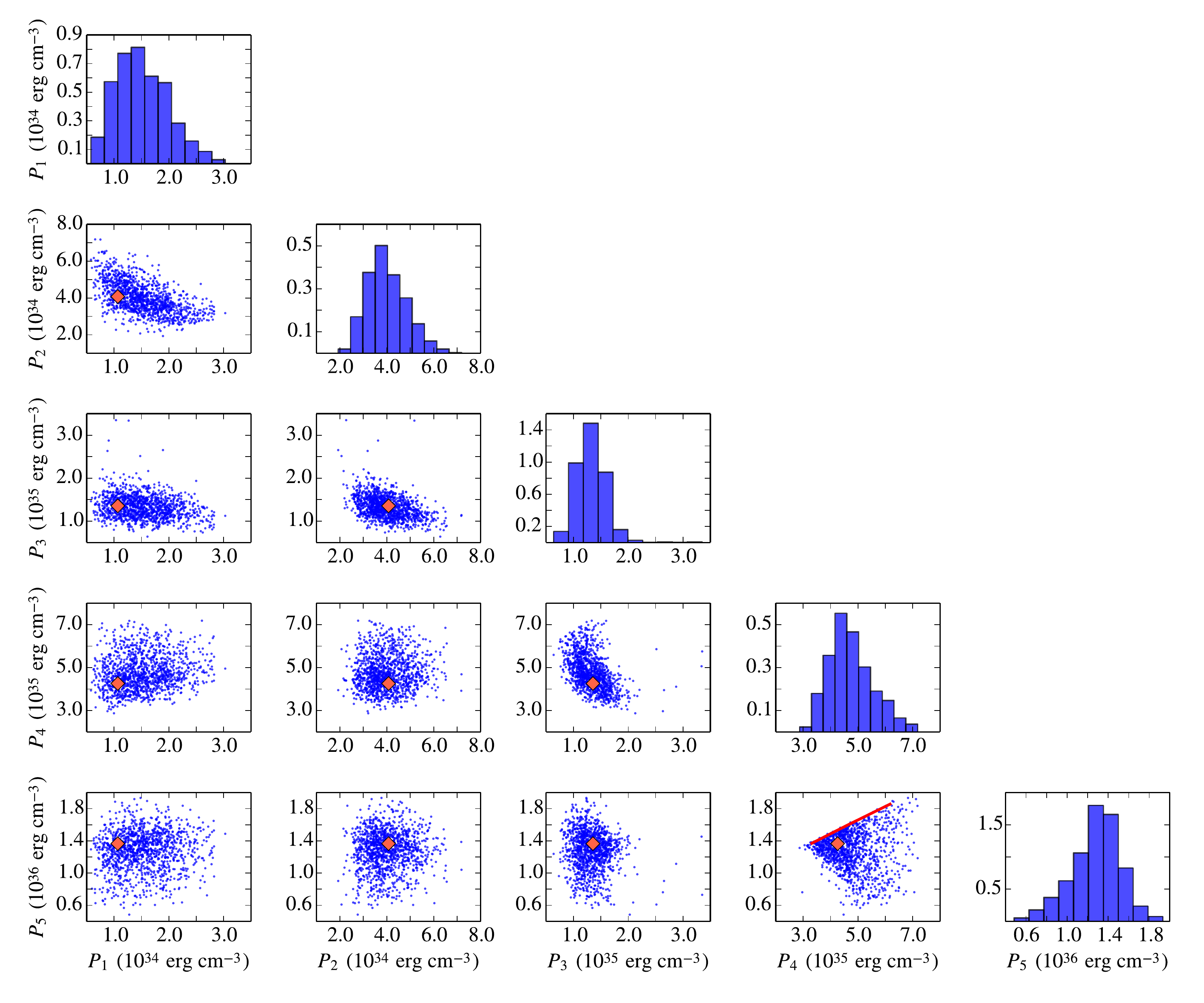}
\caption{\label{fig:corner}  Correlation plots for the inferred pressures at our five fiducial densities. The inversion was performed using uniform priors and a Gaussian regularizer ($\lambda=8$). The red diamonds mark the pressures of the true EoS (SLy) that we are trying to infer. There are slight anti-correlations between adjacent pressures (e.g., $P_1$ and $P_2$), but the non-adjacent pressures are relatively uncorrelated. The triangular shape of the $P_4$ vs. $P_5$ correlation is due to the causality requirement, which is shown as the red line.}
\end{figure*}

\section{Testing the Bayesian Inference with Mock Data}
\label{sec:data}
In this section, we test the Bayesian inference method described in $\S$\ref{sec:methods} using different sets of mock data. For most of the simulations
described below, we assume an underlying EoS and create a realization
of a sample of mass-radius data that are equidistant in mass between
1.2 and 2.0~$M_\odot$. We assume Gaussian measurement uncertainties
with the same dispersions among the data points, which we denote by
$\sigma_{\rm M}$ and $\sigma_{\rm R}$. We use a Monte Carlo method to
draw a particular realization of simulated measurements from these
distributions and apply our Bayesian inference method to this data
set. We explore below how well the Bayesian inference method works for
different types of underlying EoS, as well as for different number and
quality of data points.

In Fig.~\ref{fig:uniln}, we show the result for one realization of mock data drawn from the EoS
SLy, with uncertainties of $\sigma_{\rm R} = 0.5$~km and $\sigma_{\rm
  M}= 0.1~M_\odot$. EoS SLy is particularly challenging for a
parametrization like ours that is optimized for potentially more
complex EoS because it is practically a single polytrope in the density range
of interest. We, therefore, use this example to explore the strengths
and limits of the inference as well as of the regularizer. 

The black lines in Fig.~\ref{fig:uniln} represent the EoS SLy, while the magenta lines show the most-likely inferred EoS found with our Bayesian method. The gray bands represent the 68\% credibility regions. For five-dimensional likelihoods, the 68\% credibility region is defined as the region where
\begin{multline}
\int\int\int\int\int P(P_1, ..., P_5 | {\rm data}) dP_1 dP_2 dP_3 dP_4 dP_5 \\ = 0.68,
\end{multline}
exactly analogous to the lower-dimensional case. It should be noted that these credibility regions show the spread of possible solutions only, and should not be over-interpreted. That is, there are many curves that may be drawn through these regions that either have very small probabilities or violate one of our priors and are therefore unphysical. For example, the curve that could be drawn along the very edge of any of the bands is much less likely than those truly contained in the 68\% credibility region. To emphasize this point further, we show in Fig.~\ref{fig:indivMR} several of the individual mass-radius curves that contribute to the shape of the 68\% credibility region in the top panel of Fig.~\ref{fig:uniln}. The jagged edge of the contour is created by multiple different mass-radius curves. Indeed, the curve that follows either edge in its entirety is not included in the 68\% interval. The same can be said for all of the grayed 68\% credibility regions shown throughout this paper, and thus interpretations of those regions should be made with caution.

In the top panel of Fig.~\ref{fig:uniln}, we show the results of the inversion using uniform priors in pressure
in the absence of any regularizers. The inferred EoS contains several sharp transitions between different polytropic indices. The middle panel shows an inversion for identical data, but in which the Gaussian regularizer has been added to the uniform priors. The stepped behavior that was shown in the top panel is effectively eliminated by the regularizer. With the addition of the regularizer, our inferred, most-likely EoS closely follows SLy. The errors in pressure for our most-likely EoS are all less than 30\% for this realization of mock data, while the errors for $P_2 - P_5$ are 3$-$7\%.

Because there is no physical motivation to assume that the prior distribution of the pressures is uniform, we also tested the inversion with priors that are uniform in the logarithm of pressure. The results of this test, with identical data as above and the Gaussian regularizer included, are shown in the bottom panel of Fig.~\ref{fig:uniln}.

The top panel of Fig.~\ref{fig:uniln} shows that assuming a uniform prior introduces a preference towards higher pressures. On the other hand, assuming that prior is uniform in the logarithm of pressures biases the results toward lower pressures. The inversion is particularly sensitive to this bias in the low-mass/low-density region of the EoS, where we lack data. Our Gaussian regularizer helps reduce this bias, as shown in the middle and bottom panels of Fig.~\ref{fig:uniln}, in which the results are similar for the two types of priors when the regularizer is also included.

In the absence of a regularizer, the freedom introduced by using five polytropes in the parametrization combined with this sensitivity to prior distributions could lead to significantly skewed results, or even the false inference of a phase transition with a perceived high statistical confidence. It is, therefore, important to use a Gaussian regularizer on the second derivative, with characteristic scale $\lambda=8$, to avoid this sensitivity to over-fitting.

In Fig.~\ref{fig:corner}, we explore potential correlations between the inferred parameters for this test. Specifically, we show the correlations for the inversion using uniform priors and the Gaussian regularizer (i.e., the middle panel of Fig.~\ref{fig:uniln}).  There are slight anti-correlations between adjacent inferred pressures, due to fact that, even with the addition of the regularizer, we are still over-parametrizing our model while trying to fit the effectively single-polytrope EoS SLy. Overall, however, the pressures that are not adjacent are uncorrelated with one another. \citet{Ozel2009} showed that, for a three-polytrope parametrization, parametrizing with the pressures as free parameters instead of the polytropic indices reduced the correlations between the inferred values. Figure~\ref{fig:corner} shows that the low levels of correlations are maintained here.

\begin{figure}[ht]
\centering
\includegraphics[width=0.5\textwidth]{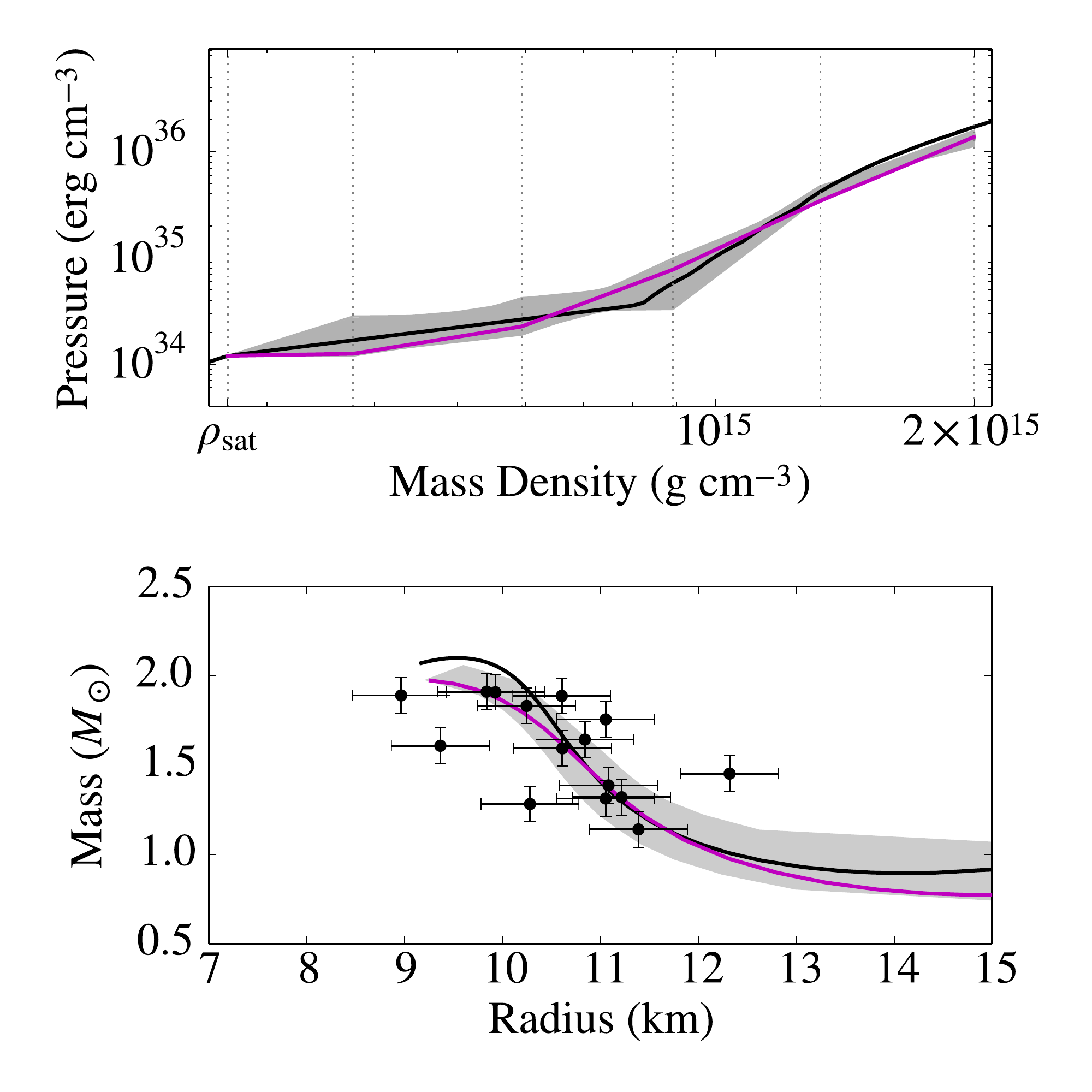}
\caption{\label{fig:EoS127}  Inferred equation of state and mass-radius curve from a sample of mock data, assuming a uniform prior distribution of pressures and a Gaussian regularizer.  The mock data are generated from a two-polytrope EoS that we created to have a break in polytropic index that does not line up with our fiducial densities, the locations of which we show with dotted vertical lines to emphasize the misalignment. The generating EoS is shown in black, our most likely inferred EoS is shown in magenta, and the gray regions represent the 68\% credibility regions.}
\end{figure}

In order to ensure that the Bayesian inference works well for other
underlying EoS and that the regularizer does not adversely limit our
ability to detect potential phase transitions, we tested the method on
a number of different EoS as well.  As an example,\footnote[1]{See discussion
  around Fig.~\ref{fig:rdiff} for other examples of EoS with a wide range of
  predicted radii.} we show in Fig.~\ref{fig:EoS127} the results of our
method obtained for an underlying EoS with significantly more
structure than SLy; specifically, we generated a mock EoS with an
extreme change in the polytropic index (from $\Gamma=1$ to $\Gamma=5$)
that occurs in between two of our fiducial densities. This EoS is shown in
the black solid line of Fig.~\ref{fig:EoS127}. This was again designed to challenge the inversion procedure, but in the opposite extreme from EoS SLy. Even in this case, the most likely solution still recovers all of the pressures to within $\sim$30\%, and recovers $P_2$ to within 11\%.

Some previous studies (e.g., \citealt{Steiner2016}) have suggested that parametrizing with polytropic segments disfavors phase transitions because polytropes naturally go through the origin. However, continuity between segments, as required in any reasonable parametrization, implies that the power law segments are never required to go through the origin. Small values of the exponent, and hence phase transitions, are thus fully allowed. Moreover, in Fig.~\ref{fig:EoS127}, we show that our parametrization is able to recover a phase transition, even when the phase transition is offset from the fiducial densities.

\begin{figure}[ht]
\centering
\includegraphics[width=0.5\textwidth]{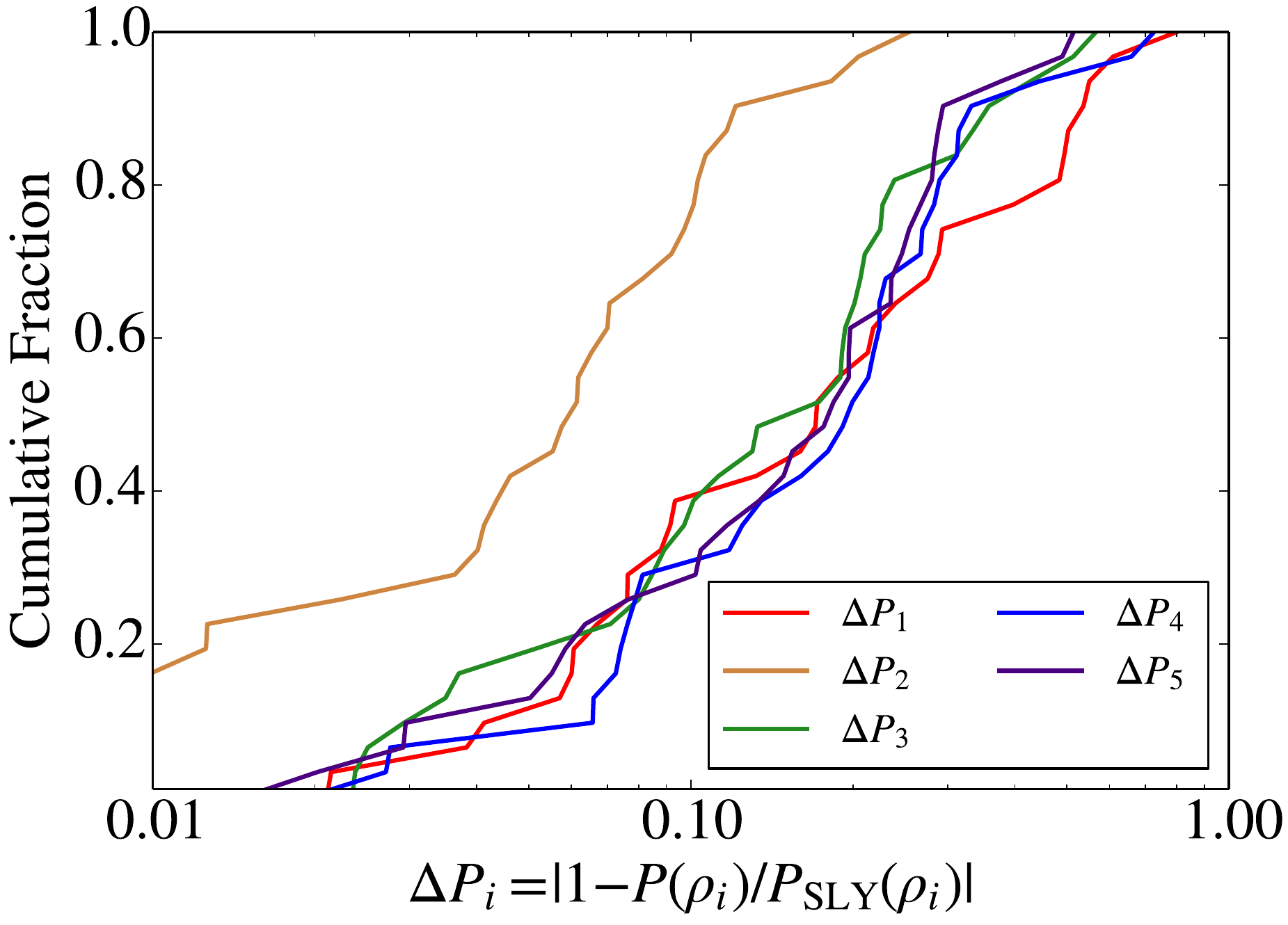}
\caption{\label{fig:Pcum}  Cumulative distribution of the errors in the most likely inferred pressure at each of the five fiducial densities for 32 realizations of mock data. For each realization, we took 15 mock $(M, R)$ data points from the EoS SLy and dithered with noise drawn from Gaussian distributions corresponding to $\sigma_M = 0.1~\Ms$ and $\sigma_R$ = 0.5~km.  We find that $P_2$ is the best constrained, with errors less than 15\% in 95\% of cases, while the other pressures have errors less than $\sim$20\% in half of the realizations and are correct to within 0.3~dex in every realization. }
\end{figure}

Finally, in order to quantify the range of uncertainties in the
inferred pressures for different statistical realizations of the mock
measurements, we generate a large number of mock
 data sets drawn from
the EoS SLy (as in Figs.~\ref{fig:uniln}$-$\ref{fig:corner}) and applied our method to each
set. We summarize these results in Fig.~\ref{fig:Pcum}, which shows the cumulative distribution of errors in the most likely inferred pressures at each of our five fiducial densities.  We find that the pressure at $\rho_2 = 2.2~\rns$ is the best constrained, with errors less than 15\% in 95\% of the realizations. The other four pressures have errors less than $\sim$20\% in approximately half of the realizations, and are correct to within less than 0.3~dex in every realization.

\begin{figure*}[ht]
\centering
\includegraphics[width=0.95\textwidth]{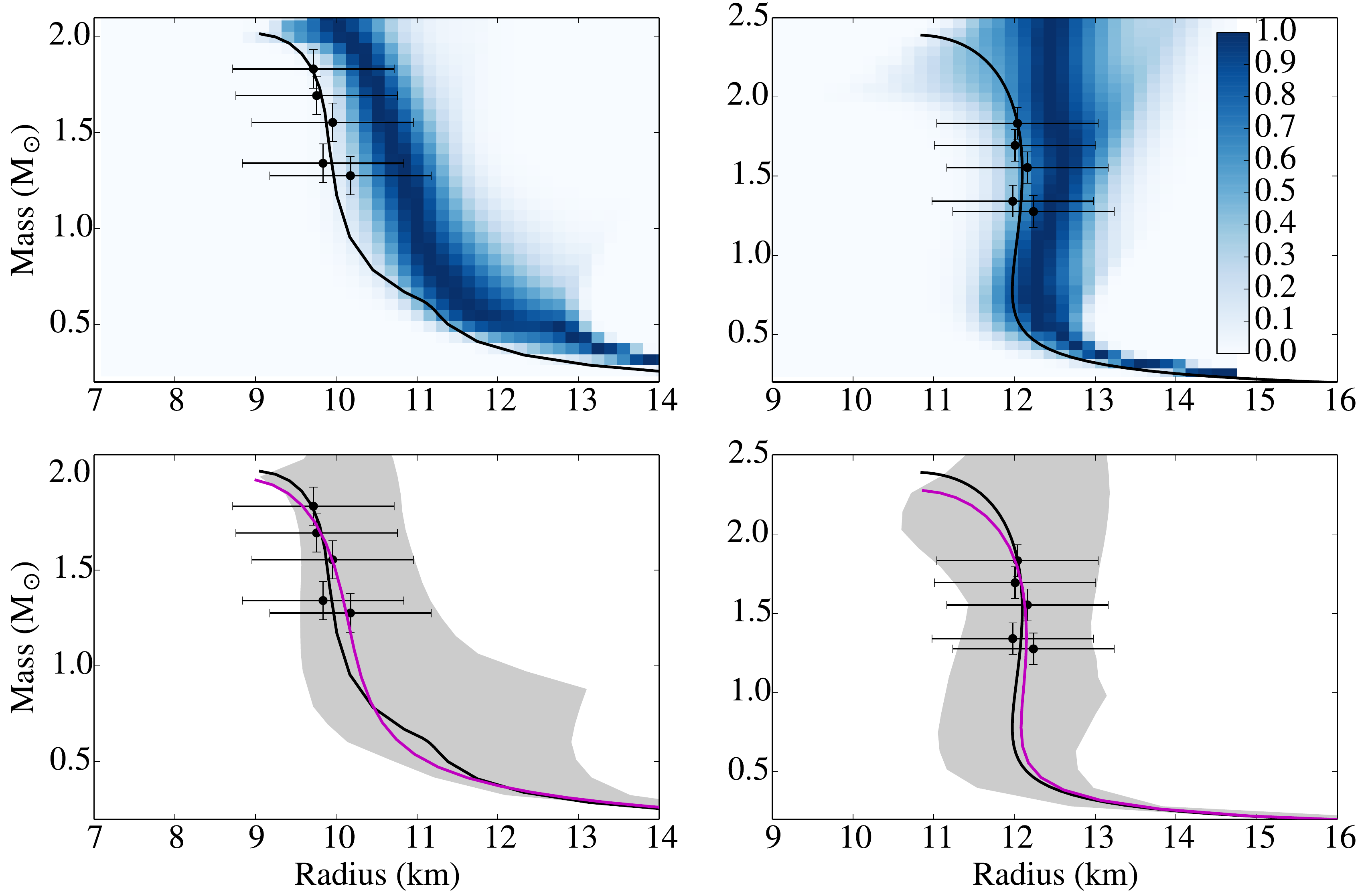}
\caption{\label{fig:marg5D}  Top panel: Marginalization of the most common radii found during the Bayesian inference for two separate EoS (left and right panels). This marginalization is computed using histograms of radii for a fixed grid of masses, for which the analytic equivalent is shown in eq.~(\ref{eq:marg}). Bottom panel: The probability distribution from the same Bayesian inference. The most likely solution is shown in magenta, while the gray regions show the 68\% credibility regions. In both cases, the mock data and generating EoS are shown in black. For the data centered near 10~km, the marginalized contours are offset from the most likely solution by $\sim$0.5~km. For the data clustered near 12~km, the marginalization bias is smaller but still present.}
\end{figure*}

\section{Biases due to marginalization}
\label{sec:marg}

The Bayesian inference scheme described in $\S$\ref{sec:methods} provides five-dimensional posteriors, $P(P_1, ..., P_5 | \mathrm{data})$. While one might explore the marginalized distributions over each pressure, $P_i$, this can easily lead to misinterpretations. It is possible, for example, that just considering the most likely pressures from each marginalized distribution of $P_i$ will produce an EoS that violates one of our priors. The pressures are coupled to one another and marginalizing over any one dimension removes that dependence. For this reason, we exclusively use the five-dimensional likelihoods to interpret our results.

Another approach that is taken in some earlier studies (e.g., \citealt{Steiner2010}) is to marginalize the output of the Bayesian inference in mass-radius space, rather than over the pressures. This is done by creating a one-dimensional histogram of the radii over a fixed grid of masses, for all possible mass-radius curves. Analytically, we can derive such a marginalization by first writing the radius and mass as functions of the five pressures and a central density, i.e., 
\begin{subequations}
\begin{equation}
R = R(\rho_c, P_1, ..., P_5)
\end{equation}
\begin{equation}
M = M(\rho_c, P_1, ..., P_5).
\end{equation}
\end{subequations}
For a fixed mass, we can write the radius equation instead as
\begin{equation}
R = R(P_1, ... P_5 ; M),
\end{equation}
which we can invert to recover $P_1$ in terms of $R$ and $(P_2, ..., P_5)$ for a fixed mass. One can then express the radius as a function of only the mass by marginalizing across the other four pressures, i.e., 
\begin{multline}
\label{eq:marg}
P(R ; M) = \int P \left[ R(P_2, ..., P_5 ; M), P_2, ..., P_5 \right]  \\ 
\times \hat{J}  \left( \frac{R}{P_1} \right) dP_2 ... dP_5,
\end{multline}
where $\hat{J} (R/P_1)$ is the Jacobian that transforms from $P_1$ to $R$. This is the analytical equivalent of taking the one-dimensional histogram of radii over a grid of masses. 

Equation~(\ref{eq:marg}) highlights the issues that marginalizing introduces. If the full posterior, $P \left[ R(P_2, ..., P_5 ; M), P_2, ..., P_5 \right]$, is relatively flat, then the highly non-linear Jacobian will dominate the resulting marginalization. The marginalization is particularly sensitive to this bias when the data are sparse and have large errors. However, the marginalization can be skewed for any data set, if the posterior distribution is not sharply peaked enough to overcome the influence of the Jacobian.

Indeed, in Fig.~\ref{fig:marg5D}, we show this effect for two different sets of data: one that is clustered around $R~\sim~10$~km and one that is clustered around $R~\sim~12$~km. Both inversions use 5 simulated $(M, R)$ data points, with masses that are spaced evenly between 1.2 and 1.8~$\Ms$ and with measurement uncertainties of $\sigma_R$=1~km and $\sigma_M = 0.1~\Ms$. For the smaller-radii dataset, the marginalized solution is offset by $\gtrsim~0.5$~km from the data at all masses (upper left panel). The most likely solution, in contrast, goes right through the data (lower left panel). For the larger-radii dataset, the effect is less extreme but still present:  the marginalized solution is shifted approximately 0.4~km to the right of the data. The most likely solution, again, goes through every data point. While the marginalized solution is indeed within the error bars of the data, the most likely solution recreates the data almost perfectly for a variety of data. Current radius data have even larger and often overlapping error contours that will make this effect hard to identify by eye. It is, therefore, extremely important that only the full five-dimensional likelihoods be used to identify EoS constraints.

It should be noted that the choice of priors does affect the size and direction of this bias. Here, we use a prior distribution that is uniform in pressure and includes the Gaussian regularizer, while still requiring our other physical constraints (e.g., causality, a 2~$\Ms$ star, etc.). Using a prior that is uniform in the logarithm of pressure pushes the bias in the other direction, i.e. toward smaller radii, and also pushes our most likely solution in that direction. The effect of the prior is stronger here than in Fig.~\ref{fig:uniln} because we have more sparse data.

\begin{figure}[ht]
\centering
\includegraphics[width=0.5\textwidth]{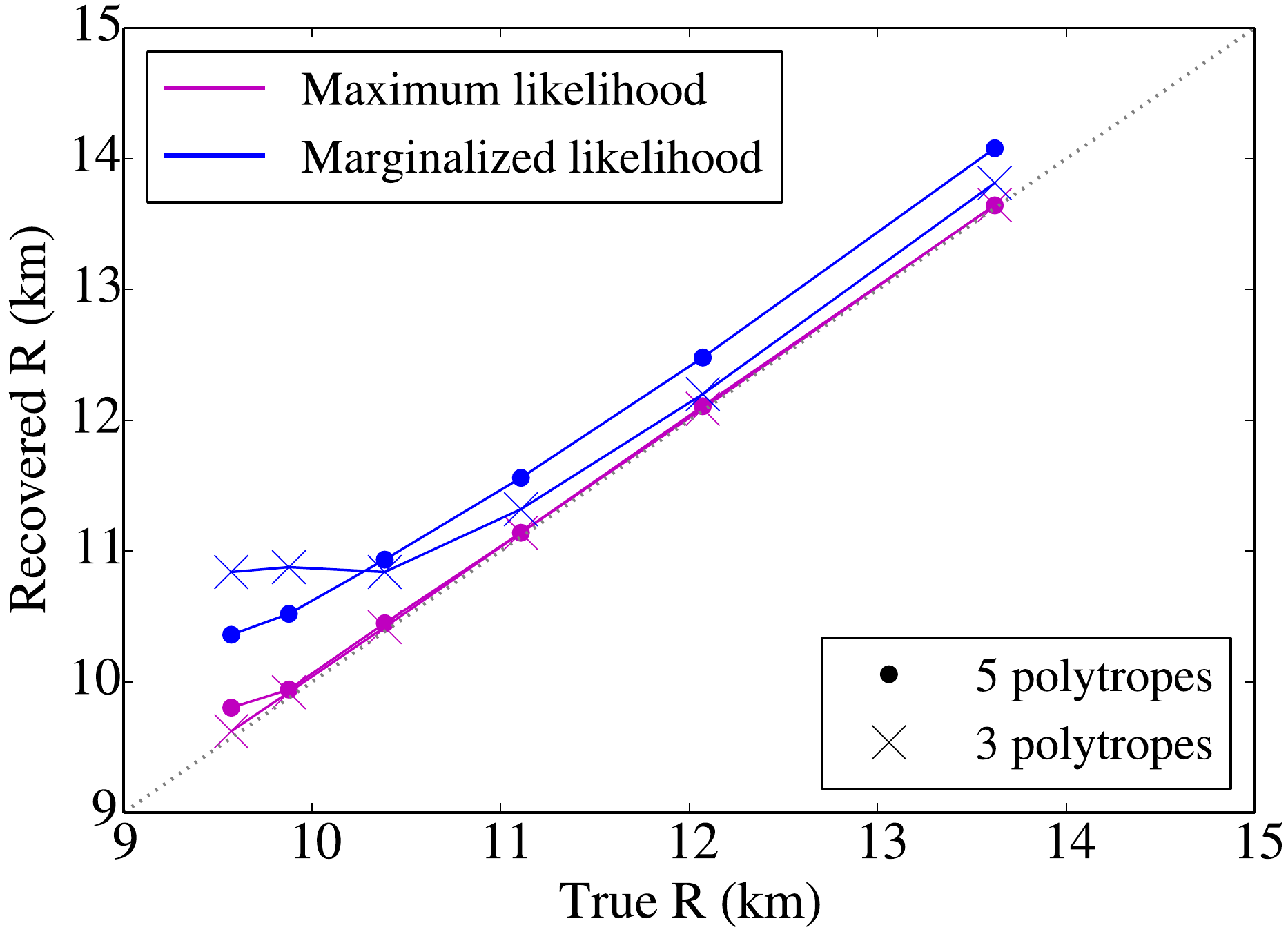}
\caption{\label{fig:rdiff}  Radius recovered in our inversion plotted against the true radius of the underlying EoS. The radii are sampled at four masses between 1.2 and 1.8~$\Ms$ and then averaged to create each value of $R$. The gray, dotted line represents a perfect inversion, with no introduced error. The magenta lines show the most likely solution, with circles indicating the solution for our five-polytrope parametrization and crosses indicating the solution for a three-polytrope parametrization. The blue lines and points represent the solutions obtained via marginalization. The most likely solution is consistent with the true value for all radii, while the marginalization solution is only correct for radii~$\ga$~12~km and only for the three-polytrope parametrization. At small radii, the marginalization biases the recovered radii to larger values by as much as  $\sim$1~km. }
\end{figure}

Figure~\ref{fig:rdiff} shows the bias of the marginalized solutions for inversions of still more data sets, each of which used uniform priors and the Gaussian regularizer. This figure also shows that the effect persists whether three or five polytropes are included in the parametrization. Each inversion used 5 simulated $(M, R)$ data points clustered near a radius between $\sim10-15$~km, with masses between $1.2-1.8~\Ms$, as in Fig.~\ref{fig:marg5D}. We find that the bias is strongest at small radii. For data centered near 9.6~$\Ms$, the bias is 1.1~km for a three-polytrope parametrization and 0.6~km for our five-polytrope parametrization. We include the results for the three-polytrope parametrization in order to emphasize that this bias is not a result of our specific choice of parametrization, but is a problem stemming from the marginalization of posteriors using sparse data. Given the large biases that can be introduced by the marginalization, it is clear that the marginalized solution should not be trusted. This is particularly true when the data have large, overlapping errors which will make this effect difficult to identify. The maximum likelihood method that we have used throughout this paper does not suffer any such biases.

\section{Conclusions}
In this paper, we developed a Bayesian method that can be used to infer the pressures of a parametrized EoS from a set of neutron star masses and radii. We used a parametrization containing five polytropic segments starting at $\rns$, which is the form we found to be optimal in our previous work \citep{Raithel2016}.

We investigated the influence of various priors and measurement uncertainties on the inferred pressures. We found that the freedom introduced by using five polytropic segments in the parametrization (which is necessary to recreate the EoS using next-generation data to within expected uncertainties) caused the inferred EoS to be too stepped in behavior. We, therefore, introduced a regularizer over the second derivative of pressure, to mitigate that freedom without sacrificing the ability to remain agnostic about the dense matter EoS. Combined with uniform priors over the pressures, we were able to show that measurement uncertainties expected in the near future will allow an inference of the pressure at $\rho_2 = 2.2\rns$ to within $\sim$15\%, for a simple EoS. We were able to recreate the pressures at $\rho_3-\rho_5$ to within $\sim$~20\% in approximately half of the realizations, and to within 0.3~dex for all realizations. For a more complicated EoS with a significant break in the power-law indices, we were able to infer the pressures at all densities to within $\sim30$\%.

Finally, we showed that determining the posterior via marginalization in mass-radius space may lead to significant biases. We found that, for data at small radii, the marginalized mass-radius curves can be biased by nearly $+1$~km. Previously published EoS inferences from neutron star radius measurements are likely to have been affected by this bias in studies that used such a marginalization. It is better, instead, to use the maximum of the five-dimensional likelihoods computed in the Bayesian inference method. The most likely solutions do not suffer any such bias and were able to recreate the true $M-R$ curve for data at any radius.

It is difficult to compare our results to previous attempts to infer the pressures of a parametrized EoS from data because other studies do not characterize the uncertainties in the same ways. For example, \citet{Steiner2016} report that the maximum uncertainty allowed in their inversion is a factor of 3 in the pressure, purely from the requirement of causality, from their assumed crust EoS, and from the requirement that an EoS produce a star with $M\geq1.97~\Ms$. They report that the maximum uncertainty is closer to a factor of 2 near the central densities of the maximum mass stars. However, these maximum uncertainties are due to the priors of their model only, and would likely be smaller with the inclusion of data. 

Using our approach, we find that the EoS can be inferred to high accuracy with the expected quality of next-generation data. Given that the EoS is currently poorly constrained at high densities, the possibility of constraining it to within even 0.3~dex, and possibly to within 15\% at 2.2~$\rns$, will allow significant advances in our understanding of the physics at work in the ultradense regime.

{\em{Acknowledgements.\/}} CR is supported by the NSF Graduate Research Fellowship Program Grant DGE-1143953. FO gratefully acknowledges a fellowship from the John Simon Guggenheim Memorial Foundation in support of this work. DP acknowledges support from the Radcliffe Institute for Advanced Study at Harvard University.

\bibliography{carolyn_2}

\begin{thebibliography}{}
\expandafter\ifx\csname natexlab\endcsname\relax\def\natexlab#1{#1}\fi

\bibitem[{{Akmal} {et~al.}(1998){Akmal}, {Pandharipande}, \&
  {Ravenhall}}]{Akmal1998}
{Akmal}, A., {Pandharipande}, V.~R., \& {Ravenhall}, D.~G. 1998, \prc, 58, 1804

\bibitem[{{Alford} {et~al.}(2005){Alford}, {Braby}, {Paris}, \&
  {Reddy}}]{Alford2005}
{Alford}, M., {Braby}, M., {Paris}, M., \& {Reddy}, S. 2005, \apj, 629, 969

\bibitem[{{Alford} {et~al.}(2013){Alford}, {Han}, \& {Prakash}}]{Alford2013}
{Alford}, M.~G., {Han}, S., \& {Prakash}, M. 2013, \prd, 88, 083013

\bibitem[{{Antoniadis} {et~al.}(2013){Antoniadis}, {Freire}, {Wex}, {Tauris},
  {Lynch}, {van Kerkwijk}, {Kramer}, {Bassa}, {Dhillon}, {Driebe}, {Hessels},
  {Kaspi}, {Kondratiev}, {Langer}, {Marsh}, {McLaughlin}, {Pennucci}, {Ransom},
  {Stairs}, {van Leeuwen}, {Verbiest}, \& {Whelan}}]{Antoniadis2013}
{Antoniadis}, J., {Freire}, P.~C.~C., {Wex}, N., {et~al.} 2013, Science, 340,
  448

\bibitem[{{Balberg} \& {Gal}(1997)}]{Balberg1997}
{Balberg}, S., \& {Gal}, A. 1997, Nuclear Physics A, 625, 435

\bibitem[{{Baym} {et~al.}(1971){Baym}, {Pethick}, \& {Sutherland}}]{Baym1971}
{Baym}, G., {Pethick}, C., \& {Sutherland}, P. 1971, \apj, 170, 299

\bibitem[{{Bogdanov} {et~al.}(2016){Bogdanov}, {Heinke}, {{\"O}zel}, \&
  {G{\"u}ver}}]{Bogdanov2016}
{Bogdanov}, S., {Heinke}, C.~O., {{\"O}zel}, F., \& {G{\"u}ver}, T. 2016, ArXiv
  e-prints, arXiv:1603.01630

\bibitem[{{Demorest} {et~al.}(2010){Demorest}, {Pennucci}, {Ransom}, {Roberts},
  \& {Hessels}}]{Demorest2010}
{Demorest}, P.~B., {Pennucci}, T., {Ransom}, S.~M., {Roberts}, M.~S.~E., \&
  {Hessels}, J.~W.~T. 2010, \nat, 467, 1081

\bibitem[{{Douchin} \& {Haensel}(2001)}]{Douchin2001}
{Douchin}, F., \& {Haensel}, P. 2001, \aap, 380, 151

\bibitem[{{Fonseca} {et~al.}(2016){Fonseca}, {Pennucci}, {Ellis}, {Stairs},
  {Nice}, {Ransom}, {Demorest}, {Arzoumanian}, {Crowter}, {Dolch}, {Ferdman},
  {Gonzalez}, {Jones}, {Jones}, {Lam}, {Levin}, {McLaughlin}, {Stovall},
  {Swiggum}, \& {Zhu}}]{Fonseca2016}
{Fonseca}, E., {Pennucci}, T.~T., {Ellis}, J.~A., {et~al.} 2016, \apj, 832, 167

\bibitem[{{Friedman} \& {Pandharipande}(1981)}]{Friedman1981}
{Friedman}, B., \& {Pandharipande}, V.~R. 1981, Nuclear Physics A, 361, 502

\bibitem[{{Gandolfi} {et~al.}(2014){Gandolfi}, {Carlson}, {Reddy}, {Steiner},
  \& {Wiringa}}]{Gandolfi2014}
{Gandolfi}, S., {Carlson}, J., {Reddy}, S., {Steiner}, A.~W., \& {Wiringa},
  R.~B. 2014, European Physical Journal A, 50, 10

\bibitem[{{Guillot} \& {Rutledge}(2014)}]{Guillot2014}
{Guillot}, S., \& {Rutledge}, R.~E. 2014, \apjl, 796, L3

\bibitem[{{Guillot} {et~al.}(2013){Guillot}, {Servillat}, {Webb}, \&
  {Rutledge}}]{Guillot2013}
{Guillot}, S., {Servillat}, M., {Webb}, N.~A., \& {Rutledge}, R.~E. 2013, \apj,
  772, 7

\bibitem[{{Hebeler} {et~al.}(2010){Hebeler}, {Lattimer}, {Pethick}, \&
  {Schwenk}}]{Hebeler2010}
{Hebeler}, K., {Lattimer}, J.~M., {Pethick}, C.~J., \& {Schwenk}, A. 2010,
  Physical Review Letters, 105, 161102

\bibitem[{{Heinke} {et~al.}(2014){Heinke}, {Cohn}, {Lugger}, {Webb}, {Ho},
  {Anderson}, {Campana}, {Bogdanov}, {Haggard}, {Cool}, \&
  {Grindlay}}]{Heinke2014}
{Heinke}, C.~O., {Cohn}, H.~N., {Lugger}, P.~M., {et~al.} 2014, \mnras, 444,
  443

\bibitem[{{Kaplan} \& {Nelson}(1986)}]{Kaplan1986}
{Kaplan}, D.~B., \& {Nelson}, A.~E. 1986, Physics Letters B, 175, 57

\bibitem[{{Kojo} {et~al.}(2015){Kojo}, {Powell}, {Song}, \& {Baym}}]{Kojo2015}
{Kojo}, T., {Powell}, P.~D., {Song}, Y., \& {Baym}, G. 2015, \prd, 91, 045003

\bibitem[{{Lattimer}(2012)}]{Lattimer2012}
{Lattimer}, J.~M. 2012, Annual Review of Nuclear and Particle Science, 62, 485

\bibitem[{{Lindblom}(1992)}]{Lindblom1992}
{Lindblom}, L. 1992, \apj, 398, 569

\bibitem[{{Lindblom} \& {Indik}(2012)}]{Lindblom2012}
{Lindblom}, L., \& {Indik}, N.~M. 2012, \prd, 86, 084003

\bibitem[{{Lindblom} \& {Indik}(2014)}]{Lindblom2014}
---. 2014, \prd, 89, 064003

\bibitem[{{N{\"a}ttil{\"a}} {et~al.}(2015){N{\"a}ttil{\"a}}, {Steiner},
  {Kajava}, {Suleimanov}, \& {Poutanen}}]{Nattila2015}
{N{\"a}ttil{\"a}}, J., {Steiner}, A.~W., {Kajava}, J.~J.~E., {Suleimanov},
  V.~F., \& {Poutanen}, J. 2015, ArXiv e-prints, arXiv:1509.06561

\bibitem[{{{\"O}zel} \& {Freire}(2016)}]{Ozel2016}
{{\"O}zel}, F., \& {Freire}, P. 2016, ArXiv e-prints, arXiv:1603.02698

\bibitem[{{{\"O}zel} \& {Psaltis}(2009)}]{Ozel2009}
{{\"O}zel}, F., \& {Psaltis}, D. 2009, \prd, 80, 103003

\bibitem[{{{\"O}zel} {et~al.}(2016){{\"O}zel}, {Psaltis}, {G{\"u}ver}, {Baym},
  {Heinke}, \& {Guillot}}]{Ozel2016a}
{{\"O}zel}, F., {Psaltis}, D., {G{\"u}ver}, T., {et~al.} 2016, \apj, 820, 28

\bibitem[{{Pandharipande} \& {Smith}(1975)}]{Pandharipande1975}
{Pandharipande}, V.~R., \& {Smith}, R.~A. 1975, Nuclear Physics A, 237, 507

\bibitem[{{Raithel} {et~al.}(2016){Raithel}, {Ozel}, \&
  {Psaltis}}]{Raithel2016}
{Raithel}, C.~A., {Ozel}, F., \& {Psaltis}, D. 2016, ArXiv e-prints,
  arXiv:1605.03591

\bibitem[{{Read} {et~al.}(2009){Read}, {Lackey}, {Owen}, \&
  {Friedman}}]{Read2009}
{Read}, J.~S., {Lackey}, B.~D., {Owen}, B.~J., \& {Friedman}, J.~L. 2009, \prd,
  79, 124032

\bibitem[{{Steiner} {et~al.}(2010){Steiner}, {Lattimer}, \&
  {Brown}}]{Steiner2010}
{Steiner}, A.~W., {Lattimer}, J.~M., \& {Brown}, E.~F. 2010, \apj, 722, 33

\bibitem[{{Steiner} {et~al.}(2016){Steiner}, {Lattimer}, \&
  {Brown}}]{Steiner2016}
---. 2016, European Physical Journal A, 52, 18

\bibitem[{{Wiringa} {et~al.}(1995){Wiringa}, {Stoks}, \&
  {Schiavilla}}]{Wiringa1995}
{Wiringa}, R.~B., {Stoks}, V.~G.~J., \& {Schiavilla}, R. 1995, \prc, 51, 38

\end{thebibliography}
\bibliographystyle{apj}

\end{document}